\documentclass[12pt]{article}

\usepackage{latexsym}
\usepackage{amsmath}
\usepackage{amssymb}
\catcode`\@=11
\newcount\hour
\newcount\minute
\newtoks\amorpm \hour=\time\divide\hour by 60\minute
=\time{\multiply\hour by 60 \global\advance\minute by-\hour}
\edef\standardtime{{\ifnum\hour<12 \global\amorpm={am}%
        \else\global\amorpm={pm}\advance\hour by-12 \fi
        \ifnum\hour=0 \hour=12 \fi
        \number\hour:\ifnum\minute<10
        0\fi\number\minute\the\amorpm}}
\edef\militarytime{\number\hour:\ifnum\minute<10
0\fi\number\minute}
\def\draftlabel#1{{\@bsphack\if@filesw {\let\thepage\relax
   \xdef\@gtempa{\write\@auxout{\string
      \newlabel{#1}{{\@currentlabel}{\thepage}}}}}\@gtempa
   \if@nobreak \ifvmode\nobreak\fi\fi\fi\@esphack}
        \gdef\@eqnlabel{#1}}
\def\@eqnlabel{}
\def\@vacuum{}
\def\marginnote#1{}
\def\draftmarginnote#1{\marginpar{\raggedright\scriptsize\tt#1}}
\def\draft{
        \pagestyle{plain}
        \overfullrule=2pt
        \oddsidemargin -.1truein
        \def\@oddhead{\sl \phantom{\today\quad\militarytime} \hfil
        \smash{\Large\sl DRAFT} \hfil \today\quad\militarytime}
        \let\@evenhead\@oddhead
        \let\label=\draftlabel
        \let\marginnote=\draftmarginnote
        \def\ps@empty{\let\@mkboth\@gobbletwo
        \def\@oddfoot{\hfil \smash{\Large\sl DRAFT} \hfil}
        \let\@evenfoot\@oddhead}
        \def\@eqnnum{(\theequation)\rlap{\kern\marginparsep\tt\@eqnlabel}%
        \global\let\@eqnlabel\@vacuum}  }

\renewcommand{\theequation}{\thesection.\arabic{equation}}
\renewcommand{\thefootnote}{\fnsymbol{footnote}}
\newcommand{\newsection}{    
\setcounter{equation}{0}\section}
\def\appendix#1{\addtocounter{section}{1}\setcounter{equation}{0}
\renewcommand{\thesection}{\Alph{section}}
\section*{Appendix \thesection\protect\indent \parbox[t]{11.15cm}{#1}}
\addcontentsline{toc}{section}{Appendix \thesection\ \ \ #1}}

\def \lc {{light-cone}}

\def \la {\label}

\def\be{\begin{equation}}
\def\ee{\end{equation}}

 \def \lc {light-cone\ }

\catcode`\@=11
\def \lc {light cone\ }

\def\bea{\begin{eqnarray}}
\def\eea{\end{eqnarray}}
\def\beann{\begin{eqnarray*}}
\def\eeann{\end{eqnarray*}}
\def\beq{\begin{equation}}
\def\eeq{\end{equation}}
\def\ba{\begin{array}}
\def\ea{\end{array}}
\def\ben{\begin{enumerate}}
\def\een{\end{enumerate}}
 \def \l {\lambda}

 \def \la {\label}
 \def\be{\begin{equation}}
\def\ee{\end{equation}}

\def \la {\label}

\font\mybb=msbm10 at 11pt

\def\bb#1{\hbox{\mybb#1}}

\def\bZ {\bb{Z}}
\def\bR {\bb{R}}

\def\bC {\bb{C}}

\def\e  {\epsilon}

\def \ee {\epsilon}

\def \l {\lambda}

\def\lc{\lrcorner}

\newcommand{\eps}{\epsilon}

\def\be{\begin{equation}}
\def\ee{\end{equation}}

\def \la{\label}

\begin{document}
\date{November 2002}
\begin{titlepage}
\begin{center}
\vspace*{3.5cm}
{\Large \bf Geometry of all supersymmetric  four-dimensional ${\cal N}=1$ supergravity  backgrounds}\\[.2cm]

\vspace{1.5cm}
 {\large  U.~Gran$^1$, J.~Gutowski$^2$ and   G.~Papadopoulos$^3$  }

\vspace{0.5cm}

${}^1$ Fundamental Physics\\
Chalmers University of Technology\\
SE-412 96 G\"oteborg, Sweden\\

\vspace{0.5cm}
${}^2$ DAMTP, Centre for Mathematical Sciences\\
University of Cambridge\\
Wilberforce Road, Cambridge, CB3 0WA, UK

\vspace{0.5cm}
${}^3$ Department of Mathematics\\
King's College London\\
Strand\\
London WC2R 2LS, UK\\

\end{center}
\vskip 1.5 cm
\begin{abstract}
We solve the Killing spinor equations of ${\cal N}=1$ supergravity, with four supercharges, coupled to any number of vector
and scalar multiplets  in all cases. We find that backgrounds with $N=1$ supersymmetry admit a null, integrable,
Killing vector field. There are two classes of $N=2$ backgrounds. The spacetime in the first class admits a parallel null vector field and
so it is a pp-wave. The spacetime of the other class admits three Killing vector fields, and a vector field
that commutes with the three Killing directions. These backgrounds
are of cohomogeneity one with  homogenous sections  either  $\bR^{2,1}$ or $AdS_3$ and  have
an interpretation as  domain walls.
 The $N=3$  backgrounds are locally  maximally
supersymmetric. There are  $N=3$ backgrounds which arise as discrete identifications of maximally supersymmetric ones.
The maximally supersymmetric backgrounds are locally isometric to either $\bR^{3,1}$ or $AdS_4$.

\end{abstract}
\end{titlepage}
\newpage
\setcounter{page}{1}
\renewcommand{\thefootnote}{\arabic{footnote}}
\setcounter{footnote}{0}

\setcounter{section}{0}
\setcounter{subsection}{0}
\newsection{Introduction}

Four-dimensional supergravity coupled to vector and scalar multiplets with ${\cal N}=1$ supersymmetry, four supercharges,   is
 a minimal supersymmetric extension  of the standard model. Because of this, it has widespread  applications
  in particle physics phenomenology. The  theory
has been developed in stages beginning from the construction of pure supergravity \cite{sugra1, sugra2}. The couplings
to the vector and scalar multiplets were added later\footnote{The theory has appeared  in the literature in many  different conventions.
We shall mostly follow those of \cite{wess}, page 212.}, see e.g.~\cite{wess} and references within.

In recent years and following the work of Paul Tod \cite{tod}, there has been much interest in the systematic understanding of
 supersymmetric configurations of supergravity theories. In lower-dimensional supergravities, the focus has been on the
 classification of supersymmetric solutions of four- and five-dimensional theories  with more than 8 supercharges, see e.g.~\cite{gutowski,ortin, roest}.
 Special
 supersymmetric solutions
 of ${\cal N}=1$ four-dimensional theories are also known. These include the stringy  cosmic strings \cite{vafa, jggp, kallosh}, domain walls
 \cite{cvetic, townsend, gutb}
 and pp-waves.

In this paper, we solve the Killing spinor equations of four-dimensional ${\cal N}=1$ supergravity coupled
to {\it any number} of vector and scalar multiplets in {\it all cases}. For this we use the spinorial geometry approach of \cite{jguggp}.
We find that there are backgrounds
with $N=1$, $N=2$, $N=3$ and $N=4$ supersymmetry. The spacetime metric of backgrounds with $N=1$ supersymmetry
admits an integrable, null, Killing vector field.  Adapting appropriate coordinates, the metric is given in
(\ref{lcmetr}) and  (\ref{frame}).
There are two kinds of $N=2$ backgrounds. One admits a parallel null, Killing vector field and the metric is that
of a pp-wave. The other admits three Killing vector fields and an additional vector field that commutes with the three
Killing ones.  The metric is given in special coordinates (\ref{metrb}). These backgrounds  are of cohomogeneity one with
homogeneous sections either $\bR^{2,1}$ or $AdS_3$.   The $N=3$
backgrounds are locally maximally supersymmetric. However, we have shown by adapting the results
of \cite{FigueroaO'Farrill:2007kb} that there are $N=3$ backgrounds which arise
as discrete quotients of maximally supersymmetric ones.  The maximally supersymmetric backgrounds are locally isometric to either
$\bR^{3,1}$ or $AdS_4$.

This paper has been organized as follows. In section two, we state the Killing spinor equations which arise
from the supersymmetry variation of the fermions of the supergravity theory. In section three, we solve the
Killing spinor equations of $N=1$ backgrounds and describe the geometry of spacetime. In section four,
we investigate the solution of the Killing spinor equations for $N=2$ backgrounds. In section five,
we show that the $N=3$ backgrounds are locally maximally supersymmetric and that the $N=4$ backgrounds
are locally isometric to  either $\bR^{3,1}$ or $AdS_4$. In section six, we give an example of an $N=3$ background which can be constructed
as discrete identification of $AdS_4$ and in section seven we give our conclusions. In appendix A, we present
the integrability conditions of the Killing spinor equations.

\newsection{Killing spinor equations}\label{killing}


The Killing spinor equations can be read off from the supersymmetry transformations of ${\cal N}=1$ supergravity.
There are three  Killing spinor equations associated with the supersymmetry transformations of the fermions
of the gravitational, gauge and scalar multiplets, respectively. After some  apparent changes in notation  from that of   \cite{wess},
the Killing spinor equations of ${\cal N}=1$ supergravity can be written as follows:

The gravitino Killing spinor equation is
\bea
\label{graveq}
2[\nabla_\mu\e_L+{1\over4} (\partial_iK\, {\cal D}_\mu \phi^i-\partial_{\bar i}K\,{\cal D}_\mu \phi^{\bar i})\e_L]+i e^{{K\over2}} W\gamma_\mu\e_R=0~,
\eea
the gaugino Killing spinor equation is
\bea
\label{gaugeq}
F^a_{\mu\nu}\gamma^{\mu\nu}\e_L-2i  \mu^a\e_L=0~,
\eea
and the Killing spinor equation associated with the scalar multiplet is
\bea
\label{mateq}
i \gamma^\mu \e_R {\cal D}_\mu \phi^i- e^{{K\over2}} G^{i\bar j} D_{\bar j} \bar W \e_L=0~,
\eea
where $\nabla$ is the spin connection, $\phi^i$ is a complex scalar  field, $K=K(\phi^i, \phi^{\bar j})$ is the K\"ahler potential
of the (K\"ahler)
 scalar or sigma model manifold $S$, $G_{i\bar j}=\partial_i \partial_{\bar j} K$, $W=W(\phi^i)$ is a (local) holomorphic function on $S$,
\bea
D_iW=\partial_iW+\partial_iK W~,~~~{\cal D}_\mu \phi^{i}=\partial_\mu\phi^i- A_\mu^a \xi^i_a~,
\eea
 $\xi_a$ are  holomorphic Killing  vector fields on $S$, $A^a$ is the gauge connection with field strength $F^a$ and
$\mu^a$ is the moment map, i.e.
\bea
G_{i\bar j} \xi_a^{\bar j}=i\partial_i\mu_a~.
\eea
We mostly follow the metric and spinor conventions of \cite{wess}. In particular, the spacetime metric has signature mostly plus,
$\e$ is a Majorana spinor and $\e_{L,R}={1\over2} (1\pm \gamma_5)\e$, where $\gamma_5^2=1$. We have set the gauge coupling to 1.

The gravitino Killing spinor equation is a parallel transport equation for a connection which, apart
from the Levi-Civita part, contains additional terms that depend on the matter couplings. The gauge group
of the Killing spinor equations is $Spin_c(3,1)=Spin(3,1)\times_{\bZ_2} U(1)$. The $Spin(3,1)$ acts on $\e$ with the Majorana representation
while $U(1)$ acts on the chiral component $\e_L$ with the standard 1-dimensional representation and on the anti-chiral $\e_R$
with its conjugate. The additional $U(1)$ gauge transformation is due to the coupling of the spinor $\e$ to the
$U(1)$ connection constructed from the K\"ahler potential $K$ associated with the matter couplings. In what follows,
we  use only the $Spin(3,1)$ component of the gauge group to choose the representatives of the Killing spinors.
Incidentally,
the holonomy of the supercovariant connection is contained in $Pin_c(3,1)$. This can be easily seen from the
expression for the integrability condition of the gravitino Killing spinor equation in appendix A. The additional $U(1)$
component in the holonomy group is again due the the K\"ahler potential coupling mentioned above.

\newsection{N=1 backgrounds}

\subsection{Killing spinor}

The starting point in the spinorial geometry approach \cite{jguggp} to solving Killing spinor equations is the choice of a normal form
for the Killing spinors up to gauge transformations. We have already mentioned that the gauge group is $Spin_c(3,1)$.
It is known that $Spin(3,1)=SL(2,\bC)$ and the chiral (Weyl) representation is identified with the standard representation of
$SL(2,\bC)$ on $\bC^2$. The Majorana representation which is relevant here is simply ${\bf 2}\oplus \bar {\bf 2}$
with $\bar {\bf 2}$ the complex conjugate of ${\bf 2}$.  Using the explicit realization  of spinors in terms of forms,
 the chiral representation is identified with
even forms, $\Lambda^{{\rm ev}}(\bC^2)$,  and the anti-chiral with odd ones, $\Lambda^{{\rm odd}}(\bC^2)$. Introducing
a Hermitian basis $(e_1, e_2)$ in $\bC^2$ with respect to a Hermitian  inner product $<\cdot, \cdot>$, a basis in $\Lambda^{{\rm ev}}(\bC^2)$
is $(1, e_{12})$, $e_{12}= e_1\wedge e_2$, and a basis in $\Lambda^{{\rm odd}}(\bC^2)$ is $(e_1, e_2)$.
In particular, the gamma matrices act on $\Lambda^{{\rm ev}}(\bC^2)$ and $\Lambda^{{\rm odd}}(\bC^2)$ as
\bea
\Gamma_0&=&-e_2\wedge+e_2\lc~,~~~\Gamma_2=e_2\wedge+e_2\lc~,
\cr
\Gamma_1&=&e_1\wedge+e_1\lc~,~~~\Gamma_3=i(e_1\wedge-e_1\lc)~,
\la{bas}
\eea
where $\lc$ is the adjoint operation of the form skew-product.
For later use, we also adopt a light-cone Hermitian basis in the space of spinors as
\bea
\gamma_+ &=& \sqrt{2} \,e_2 \lc~,~~~ \gamma_- = \sqrt{2}\, e_2 \wedge~,
\cr
\gamma_1 &=& \sqrt{2}\, e_1 \wedge~,~~~\gamma_{\bar{1}} = \sqrt{2}\, e_1 \lc~.
\eea
There is one orbit of $SL(2,\bC)$ on $\Lambda^{{\rm ev}}(\bC^2)$, and so the chiral
component of $\e$ can be chosen as $1$. In this basis, the Majorana inner product is given by
\bea
B(\eta_1, \eta_2)=<\Gamma_{12}\, \eta^*_1, \eta_2>~,
\eea
where $<\cdot,\cdot>$ is the Hermitian inner product on $\bC^2$ extended on  $\Lambda^\star(\bC^2)$, and $\eta_1, \eta_2\in \Lambda^\star(\bC^2)$.
Observe that $B$ is a bi-linear. The spacetime forms constructed
as spinor bi-linears are defined as
\bea
\tau_{\mu_1\dots \mu_k}=B(\eta_1, \gamma_{\mu_1\dots\mu_k}\eta_2)~,~~~k=0,1 \dots, 4~.
\la{bil}
\eea
The Dirac inner product in the (\ref{bas}) basis is $D(\eta_1, \eta_2)=<\Gamma_0\eta_1, \eta_2>$. Equating the Dirac and Majorana conjugates, one finds that the complex conjugation operation
is imposed by the anti-linear map,  $C=-\Gamma_{012}*$, $C^2=1$. Applying this to the spinor $1$, one finds that a   Majorana
representative for the orbit is
\bea
\e=1+e_1~,~~~~\e_L=1~,~~~\e_R=e_1~.
\la{fks}
\eea
This can be chosen as the first Killing spinor of the theory. The isotropy group of the spinor $1$ in $SL(2,\bC)$ is $\bC$.  This will be used
later to choose the second Killing spinor.

\subsection{Solution to the Killing spinor equations}

Evaluating the gravitino equation on the Killing spinor $\epsilon=1+e_1$,  we find that

\bea
\label{n1graveq}
- \Omega_{+,+-} + \Omega_{+, 1 \bar{1}} + {1 \over 2}
(\partial_iK\, {\cal D}_+ \phi^i-\partial_{\bar i}K\,{\cal D}_+ \phi^{\bar i}) &=& 0~,
\cr
\Omega_{+,+1} &=& 0~,
\cr
- \Omega_{-,+-} + \Omega_{-,1 \bar{1}} + {1 \over 2}
(\partial_iK\, {\cal D}_- \phi^i-\partial_{\bar i}K\,{\cal D}_- \phi^{\bar i}) &=&0~,
\cr
2 \Omega_{-,+ \bar{1}} + \sqrt{2} i e^{K \over 2}  W &=&0
\cr
- \Omega_{1,+-} + \Omega_{1,1 \bar{1}} + {1 \over 2} (\partial_iK\, {\cal D}_1 \phi^i-\partial_{\bar i}K\,{\cal D}_1 \phi^{\bar i}) &=&0~,
\cr
\Omega_{1,+ \bar{1}}=\Omega_{\bar{1},+ \bar{1}} &=&0~,
\cr
- \Omega_{\bar{1},+-}+ \Omega_{\bar{1}, 1 \bar{1}} +
{1 \over 2}(\partial_iK\, {\cal D}_{\bar{1}} \phi^i-\partial_{\bar i}K\,{\cal D}_{\bar{1}} \phi^{\bar i})
+ \sqrt{2} i e^{K \over 2} W &=&0~,
\eea
where $\Omega$ is the spin connection of the four-dimensional spacetime metric.

The gaugino equation ({\ref{gaugeq}}) acting on $1+e_1$ gives
\be
F^a_{+ 1} =F^a_{+-}=0, \qquad   F^a_{1 \bar{1}} -i \mu^{(a)}=0~,
\la{n1gsol}
\ee
and similarly the Killing spinor equation of the scalar multiplet  ({\ref{mateq}}) gives
\be
{\cal D}_+ \phi^i =0, \qquad \sqrt{2} i {\cal{D}}_1 \phi^i = e^{K \over 2} G^{i \bar{j}} D_{\bar{j}} {\bar{W}}~.
\la{n1msol}
\ee
The equations (\ref{n1graveq})-(\ref{n1msol}) is the linear system associated with the $N=1$ supersymmetric backgrounds.

To solve the linear system, substitute ${\cal D}_+ \phi^i =0$
into ({\ref{n1graveq}}) to find that the gravitino equations can be rewritten as
\be
\label{auxgrav1}
\Omega_{+,+-} = \Omega_{+,1 \bar{1}} =
\Omega_{+,+1} = \Omega_{-,-+} = \Omega_{1,+ \bar{1}}
= \Omega_{1,+1} = \Omega_{-,+1}+\Omega_{1,+-}=0~,
\ee
and
\bea
\Omega_{-,1 \bar{1}} + {1 \over 2} (\partial_iK\, {\cal D}_- \phi^i-\partial_{\bar i}K\,{\cal D}_- \phi^{\bar i})
&=& 0~,
\cr
i\sqrt{2}e^{{K \over 2}} W+2\Omega_{-,+ \bar{1}}&=&0~,
\cr
\Omega_{-,+1} + \Omega_{1,1 \bar{1}}
+{1 \over 2}  (\partial_iK\, {\cal D}_1 \phi^i-\partial_{\bar i}K\,{\cal D}_1 \phi^{\bar i}) &=& 0~.
\la{n1geomb}
\eea
In what follows, we explore the consequences of the above conditions on the geometry of spacetime.

\subsection{Geometry}
To proceed, write the metric in a light-cone Hermitian frame as
\bea
ds^2= 2 {\bf{e}}^- {\bf{e}}^++2 {\bf {e}}^1 {{\bf e}}^{\bar 1}~.
\la{lcmetr}
\eea
The spacetime form bilinears associated with the Killing spinor, see (\ref{bil}), are
proportional to ${\bf{e}}^-$ and ${\bf{e}}^-\wedge ({\bf{e}}^1$+${\bf{e}}^{\bar 1})$, and their
spacetime duals.
Setting ${\bf{e}}^-=X_\mu dy^\mu$, it is easy to see that
 ({\ref{auxgrav1}}) implies that
\be
\nabla_{(\mu} X_{\nu)} =0,  \qquad {\bf{e}}^{-} \wedge {d\bf{e}}^{-} =0,\qquad  {\bf{e}}^{-}\wedge {\bf{e}}^{\bar{1}} \wedge d {\bf{e}}^1 =0 \ .
\la{geomcon}
\ee
Observe also that ${\bf{e}}^{-}\wedge {\bf{e}}^1 \wedge d {\bf{e}}^1 =0$.

The first condition in (\ref{geomcon}) implies that the metric admits a null Killing vector field.
While the second implies that the distribution defined by $X$ is integrable.
As a result the metric can be written as in (\ref{lcmetr}) with
\bea
{\bf{e}}^-=H du~,~~~{\bf{e}}^+ = dv + V du + w_i dx^i~,~~~{\bf {e}}^1=\beta_1 dx^1+\beta_2 dx^2~,
\la{frame}
\eea
where $u,v,x^i$, $i=1,2$, are real coordinates,  $H,V,w_i$ are real spacetime functions independent of $v$ and $\beta_1, \beta_2$ are complex
spacetime functions. Substituting these into the last condition in (\ref{geomcon}), we find that the frame ${\bf e}^1$ and so its complex conjugate
${\bf e}^{\bar 1}$
can be chosen independent of $v$.

In fact, the basis given in ({\ref{frame}}) can be simplified further; one can work
in a gauge for which $w_1=w_2=0$ in ${\bf{e}}^+$.  To see how such a gauge may be chosen, consider the $Spin(3,1)$ gauge transformation generated
by $R \gamma_{+1} + {\bar{R}} \gamma_{+ \bar{1}}$ for $R \in \bC$, which leaves invariant $1+e_1$.
It is straightforward to see that this
gauge transformation corresponds to the following change of basis
\bea
{\bf{e}}^- & \rightarrow & {\bf{e}}^-
 \cr
{\bf{e}}^+ & \rightarrow & {\bf{e}}^+ -4|R|^2 {\bf{e}}^- -2 {\bar{R}} {\bf{e}}^1 -2 R {\bf{e}}^{\bar{1}}
 \cr
{\bf{e}}^1 & \rightarrow & {\bf{e}}^1 +2R {\bf{e}}^-
 \cr
{\bf{e}}^{\bar{1}} & \rightarrow & {\bf{e}}^{\bar{1}} +2 {\bar{R}} {\bf{e}}^- \ .
\eea
By making such a gauge transformation, one can set $w_1=w_2=0$ in ${\bf{e}}^+$.
Finally, a co-ordinate transformation in $x^1$, $x^2$ can be used to
eliminate the $du$ term from ${\bf{e}}^1$.
The basis is then given by  ({\ref{frame}}), with $w_1=w_2=0$.

The last two conditions in (\ref{n1geomb}) can be rewritten as
\bea
\sqrt{2} e^{{K\over2}} W {\bf e}^--\star ({\bf e}^1\wedge d{\bf e}^-)=0~,
\cr
\star d({\bf e}^-\wedge {\bf e}^{\bar 1})-{1\over \sqrt{2}} e^{{K\over2}} \bar W {\bf e}^--
{i\over 2} (\partial_iK\, {\cal D}_1 \phi^i-\partial_{\bar i}K\,{\cal D}_1 \phi^{\bar i}){\bf e}^- &=& 0~,
\eea
where the orientation of the spacetime is chosen as $\epsilon_{-+1\bar 1}=-i$. The first condition in (\ref{n1geomb})
cannot be written in a more covariant form. However, if one takes the fields to be independent of $u$, then the connection
part vanishes.

To solve (\ref{n1gsol}), one can locally  always choose the gauge $A_+^a=0$. The first two conditions in (\ref{n1gsol}) will then imply that
the remaining components of $A$ are independent of $v$. There is no general procedure to give an explicit solution
for the last condition (\ref{n1gsol}).

Next turn into the conditions (\ref{n1msol}) that arise from the Killing spinor equations of the matter multiplet.
In the gauge $A_+^a=0$, the first condition in (\ref{n1msol}) implies that the scalar fields can be taken
to be independent of $v$, $\partial_v\phi=0$. The last condition in (\ref{n1msol}) can be interpreted as a
holomorphic flow equation. The construction of explicit solutions will depend on the form of the K\"ahler potential
and $W$, and so of the details of the model.

\newsection{N=2 backgrounds}

\subsection{Killing spinors}

The first Killing spinor is the same as that of the $N=1$ case investigated above. So we set $\e_1=\e$, where $\e$ is given in (\ref{fks}).
To choose the second Killing spinor, consider the most general Majorana spinor
\bea
\e_2= a 1+be_{12}+ C (a 1+be_{12})~,~~~a,b\in \bC~.
\eea
The isotropy group of  $\e_1$  in $Spin(3,1)$ is  $\bC$. This  can be used to simplify the
expression for $\e_2$.
There are two cases to consider.
If $b=0$, the $\bC$ isotropy transformation leaves $\e_2$ invariant and
\bea
\e_2=a 1+ \bar a e_1~.
\la{sksa}
\eea
Linear independence of $\e_1$ and $\e_2$ requires that ${\rm Im}\, a\not=0$.

Next suppose that  $b\not=0$.  After a $\bC$ transformation with parameter $\lambda$,  one has
\bea
\e_2'= (a+\lambda b) 1+ b e_{12}+C[(a+\lambda b) 1+ b e_{12}]~.
\eea
Setting  $\lambda=-{a\over b}$, one can choose the normal form of $\e_2$ as
\bea
\e_2=b e_{12}-\bar b e_2~.
\la{sksb}
\eea
So the second Killing spinor can be chosen either as in (\ref{sksa}) or as in (\ref{sksb}) with $a,b$ promoted
to complex spacetime functions.

\subsection{Solution to the Killing spinor equations}

\subsubsection{$\e_2=a 1+ \bar a e_1$}
Consider first the case for which $\e_2=a 1+ \bar a e_1$. The linear system is easy to read off from that of the $N=1$ case.
In particular, the supercovariant connection along the $-$ light-cone direction gives
\bea
2a \Omega_{-,+\bar 1}+i \sqrt{ 2} \bar a e^{{K\over2}} W=0~.
\eea
Comparing this condition with those of the $N=1$ case, one concludes that either $W=0$ on the field configurations $\phi$ of the solution\footnote{
This does not imply that $W$ vanishes. It means that $W$ vanishes on the solution for $\phi$.}
or $a=\bar a$. If the latter is the
case, then it turns out that $a$ is also constant and so $\e_2$ is not linearly independent from $\e_1$.
It remains to choose $W=0$. In such a case, one finds that the parameter $a$ is constant, i.e.~$a\in \bC$, and the additional conditions to those of $N=1$ are
\bea
\Omega_{-,+1}=0~,~~~{\cal D}_1\phi^i=0~,~~~W=0~.
\eea
Combining these with those of $N=1$ backgrounds, we find that the gravitino and matter Killing spinor equations give
\bea
\label{auxgrav2}
\Omega_{+,+-} = \Omega_{+,1 \bar{1}} =
\Omega_{+,+1} = \Omega_{-,-+} = \Omega_{1,+ \bar{1}}
= \Omega_{1,+1} = \Omega_{-,+1}=\Omega_{1,+-}=0~,
\eea
and
\bea
\Omega_{-,1 \bar{1}} + {1 \over 2} (\partial_iK\, {\cal D}_- \phi^i-\partial_{\bar i}K\,{\cal D}_- \phi^{\bar i})= 0~,~~~\Omega_{1,1 \bar{1}}
-{1 \over 2} \partial_{\bar i}K\,{\cal D}_1 \phi^{\bar i} = 0~,
\cr W=\partial_jW=0~,~~~{\cal D}_1\phi^i={\cal D}_+\phi^i=0~.
\la{n2pp}
\eea
There are no additional conditions that arise from the gaugino Killing spinor equation apart from those that we have found
in the $N=1$ case (\ref{n1gsol}).

\subsubsection{$\e_2=b e_{12}-\bar b e_2$}

Next consider the case where $\e_2=b e_{12}-\bar b e_2$. The gravitino Killing spinor equation gives
\bea
\partial_+b=0~,~~~~ b \Omega_{+,-1}+ \bar b \Omega_{-,+\bar 1}=0~,
\cr
\partial_-b-\Omega_{-,1\bar1} b=0~,~~~\Omega_{-,-1}=0~,
\cr
\partial_1 b-b(\Omega_{1,-+}+\Omega_{+,-1}+\Omega_{1,1\bar1})=0~,~~~\Omega_{1,-1}=0~,
\cr
\partial_{\bar 1}b-b\Omega_{\bar1,1\bar1}=0~,~~~\Omega_{\bar1, -1}=0~,
\la{n2dsol}
\eea
where we have used the $N=1$ relations to simplify the expressions.
Moreover the gaugino Killing spinor equation gives
\bea
F^a_{-1}=0~,~~~~F^a_{1\bar 1}+i \mu^a=0~.
\la{n2gsol}
\eea
In addition, the Killing spinor equation associated with the matter multiplet gives
\bea
{\cal D}_-\phi^i=0~,~~~i\sqrt{2}\,\, \bar b \,\,{\cal D}_{\bar 1} \phi^i+b e^{{K\over2}} G^{i\bar j}D_{\bar j}\bar W=0~.
\la{n2msol}
\eea

\subsection{Geometry}

\subsubsection{$\e_2=a 1+ \bar a e_1$}

The geometric constraints  (\ref{auxgrav2}) imply that
 $X={\bf{e}}^-$ is covariantly constant with respect to the Levi-Civita connection. So the spacetime admits
 a parallel null Killing vector field. Such a spacetime has an interpretation as a pp-wave. Note, however, that
 the cosmic string solutions \cite{vafa} and their generalizations \cite{jggp, kallosh} also admit a null parallel vector field
 and so belong to this class
 of solutions.  In particular,  one can
choose co-ordinates $v,u$ such that $X={\partial \over \partial v}$
is a Killing vector, and ${\bf{e}}^- = du$, i.e.~the frame can be chosen as in (\ref{frame}) with $H=1$.
We have used the same symbol $X$ to denote the one-form  and the dual vector field.

The investigation of remaining conditions is similar to
that of the $N=1$ case. In particular the first condition in (\ref{n2pp}) does not have a
straightforward interpretation unless one takes the fields to be independent of $u$. In such a case
the connection term vanishes. The second condition in (\ref{n2pp})
can be written as
\bea
\star d(e^-\wedge e^1)-
{i\over 2} (\partial_iK\, {\cal D}_1 \phi^i-\partial_{\bar i}K\,{\cal D}_1 \phi^{\bar i}){\bf e}^- = 0~.
\eea
The conditions on $W$ in (\ref{n2pp}) imply that the solution for the scalars should be chosen such that
the superpotential $W$ and its first derivative vanish.

The restrictions on $\phi$ in (\ref{n2pp}) can be interpreted as light-cone pseudo-holomorphicity
conditions. However notice that the light-cone almost-hermitian distribution $({\bf e}^+, {\bf e}^1)$
is not integrable in general. However if one takes the fields to be independent of $u$, $({\bf e}^+, {\bf e}^1)$
is integrable and ${\cal D}_+\phi^i={\cal D}_1\phi^i=0$ are light-cone holomorphicity conditions. Moreover
in such a case, one can always choose a gauge locally such that $A_+^a=A_1^a=0$, since $F_{+1}=0$, and so
write $\partial_+\phi^i=\partial_1\phi^i=0$.

\subsubsection{$\e_2=b e_{12}-\bar b e_2$}

To analyze the conditions (\ref{n2dsol}) which arise from the Killing spinor equations in this case,
it is convenient to define the 1-forms
\be
X= {\bf{e}}^-, \qquad Y= |b|^2 {\bf{e}}^+, \qquad Z = {\bar{b}} {\bf{e}}^1 + b {\bf{e}}^{\bar{1}},
\qquad W =  i {\bar{b}} {\bf{e}}^1 -i b {\bf{e}}^{\bar{1}} \ .
\la{frem}
\ee
Observe that $Z$ is orthogonal to $X, Y, W$, and $W$ is orthogonal to $X, Y, Z$.
Then it is straightforward to show that the Killing spinor equations imply that
$X$, $Y$ and $Z$ are all Killing vectors. Furthermore, $W$ is closed, $d W =0$.
In addition, one finds the following constraints on the vector field commutators:
\be
[W, X] = [W, Y] = [W, Z] =0
\ee
and
\be
[X,Y] = c Z, \qquad [X, Z]=-2c X, \qquad [Y, Z]=2cY~,
\ee
where $c=b(\Omega_{-,+1}-\Omega_{+,-1})$ and we use the same symbols to denote the vector fields their dual
one-forms.

Consider the commutator $[X,Y] = c Z$. Since $W$ commutes with the other three vector field, the Jacobi identity
implies that $Wc=0$. Similarly, the Jacobi identity for $Z, X$ and $Y$ together with the linear independence
of these three vector field imply that $Xc=Yc=Zc=0$. So $c$ can be taken to be a constant.

Next, since $Z$ and $W$ commute one can choose coordinates $x, y$ such that $Z=\partial_x$ and $W=\partial_y$. Moreover,
the rest of the commutators imply that there are additional coordinates  $u$, $v$ such that
\be
X= e^{2cx} \partial_v , \qquad Y= e^{-2cx}\bigg((c^2 v^2+2c\lambda(u)v+\nu(u)) \partial_v
+(cv+\lambda(u)) \partial_x + \rho(u)\partial_y+ \partial_u  \bigg) \ ,
\la{vf}
\ee
where $\lambda, \nu$ and $\rho$ are arbitrary functions of $u$. The functions $\lambda$ and $\rho$ can be eliminated
using a $u$-depedent shift transformation in $v$ and $y$. The resulting expression for $Y$ is as in (\ref{vf}) with $\l=\rho=0$.
The rest of the vector fields remain unchanged.
Using (\ref{frem}),  one can compute the frame
in terms of the coordinates $x,y,v,u$ to find
\bea
{\bf e}^-&=&e^{2cx} |b|^2 du~,~~~{\bf e}^+= e^{-2cx} (dv-(c^2 v^2+\nu) du)~,~~~
\cr
{\bf e}^{1}&=&b [(dx-idy)- cv du]~,~~
{\bf e}^{\bar 1}=\bar b [(dx+idy)- cv du]~.
\eea

Hence the spacetime metric can be written as
\bea
ds^2=2|b|^2 [ds^2(M_3)+dy^2]~,
\la{metrb}
\eea
where
\be
ds^2(M_3) = du (dv-(c^2 v^2+\nu) du)+(dx-cv du)^2 \ ,
\ee
and $\nu$ is a function of $u$, $\nu=\nu(u)$.
However, by direct examination of the Riemann curvature tensor, we find that
the 3-manifold with metric $ds^2(M_3)$ is either $\bR^{2,1}$ if $c=0$, or $AdS_3$ if $c \neq 0$.

The function $b$ depends only on $y$, satisfying
\be
{d b \over dy} = \sqrt{2} |b|^2 e^{{K \over 2}} W
+{1 \over \sqrt{2}} e^{{K \over 2}} b \big( b \partial_i K G^{i \bar{j}}
D_{\bar{j}} \bar{W} - \bar{b} \partial_{\bar{i}} K G^{\bar{i} j} D_j W \big) \ .
\la{f1}
\ee
If $b$ is taken to be real, the above equation can be further simplified to write
\bea
{d \log b \over dy} = \sqrt{2}  e^{{K \over 2}} {\rm Re}\, W~,~~~ i {\rm Im}\, W+{1\over 2} \big(  \partial_i K G^{i \bar{j}}
D_{\bar{j}} \bar{W} -  \partial_{\bar{i}} K G^{\bar{i} j} D_j W \big) =0 \ .
\eea
Clearly, the spacetime is of cohomogeneity one with homogenous section either $AdS_3$ or $\bR^{2,1}$. So
this class of $N=2$ solutions can be thought of as domain wall spacetimes.

The gaugino Killing spinor equation implies that
\be
F^a =0, \qquad \mu^a =0~.
\ee
So the gauge connection is flat and can locally be set to zero. The vanishing of the moment map restricts the
scalars to lie on a K\"ahler quotient of $S$.

The scalars $\phi^i$ are independent of $v$. Since we have set $A=0$ locally,
the additional constraints on ${\cal{D}} \phi^i$
imply that $\partial_x\phi^i=\partial_u\phi^i=0$.
Moreover, the remaining Killing spinor equations of the scalar multiplet (\ref{n2msol})  gives
\be
{d \phi^i \over dy} = -\sqrt{2}\, b e^{{K \over 2}} G^{i \bar{j}} D_{\bar{j}} {\bar{W}}~.
\la{f2}
\ee
Observe that this expression depends on $b$. This is again a flow equation driven by the holomorphic potential $W$.
One can change parameterisation to simplify the flow equations (\ref{f1}) and (\ref{f2}). The construction of
explicit solutions depends on the details of the models.

\newsection{N=3 and N=4 backgrounds}

\subsection{Killing spinors}

To find the Killing spinors of $N=3$ backgrounds, we use the gauge group to bring the normal to the Killing spinors to a canonical form
as in  \cite{preons}. Since
there is a single orbit of $Spin(3,1)$ on the space of Majorana spinors, we can always choose the normal direction to the three Killing spinors to
 be  $i(e_2+e_{12})$ with respect
 to the Majorana inner product, $A(\zeta, \eta)=<\Gamma_{12}\zeta^*, \eta>$,
where $<,>$ is the standard Hermitian inner product. The orthogonal directions to $i(e_2+e_{12})$ are
 $\{\eta_r\}=\{1+e_1, e_2-e_{12}, i(e_2+e_{12})\}$. So
the three Killing spinors can be chosen as
\bea
\e_r=\sum_s f_{rs} \eta_s~,~~~r,s=1,2,3~,
\la{n3ks}
\eea
where $(f_{rs})$ is a real $3\times 3$ invertible matrix of spacetime functions. Schematically we write $\e=f\eta$.

In the $N=4$ backgrounds, the Killing spinors can again be written as a linear combination of the basis $\{1+e_1, i(1-e_1), e_2-e_{12}, i(e_2+e_{12})\}$
of Majorana spinors with real spacetime functions as coefficients. Next we shall solve the Killing spinor equations for both cases.

\subsection{Solution to the Killing spinor equations}

Let us begin with the $N=3$ case. We shall first solve the Killing spinor equations locally.
To proceed observe that (\ref{n3ks}) implies that schematically $\e_L=f \eta_L$ and $\e_R=f \eta_R$. Substituting this
into the  gaugino (\ref{gaugeq}) and chiral (\ref{mateq}) Killing spinor equations, one finds
that the
dependence on $(f)$ can be eliminated, because $f$ is invertible.
Moreover the conditions that one obtains  are those of (\ref{n1gsol}), (\ref{n1msol}), and
(\ref{n2gsol}) and (\ref{n2msol}) for $b=1$ and $b=i$. These  imply that
\bea
F^a_{\mu\nu}={\cal D}_\mu\phi^i=D_i W=\mu^a=0~.
\la{n3sol}
\eea
Since the gauge connection is flat, we can locally set the gauge potential to vanish, $A^a_\mu=0$. As a result the second equation implies that
$\phi$ are constant. Substituting these data into the gravitino Killing spinor equation, and taking its integrability condition,
we find that
\bea
R_{\mu\nu, \rho\sigma} \gamma^{\rho\sigma}\eta_L+2 e^K W\bar W \gamma_{\mu\nu}\eta_L=0~.
\la{n3int1}
\eea
Clearly the integrability condition takes values in $\mathfrak{spin}(3,1)$. Since the isotropy group of three linearly independent spinors
in $Spin(3,1)$ is the identity, (\ref{n3int1}) implies that
\bea
R_{\mu\nu, \rho\sigma}=- e^K W\bar W (g_{\mu\rho} g_{\nu\sigma}-g_{\mu\sigma} g_{\nu\rho})~.
\la{n3int}
\eea
It is easy to see that (\ref{n3sol}) and (\ref{n3int}) are precisely the conditions that one gets for backgrounds that admit $N=4$
supersymmetries. So one concludes that $N=3$ backgrounds admit locally an additional
supersymmetry and so are locally maximally supersymmetric. Furthermore (\ref{n3int}) implies that the spacetime
is either $\bR^{3,1}$ or $AdS_4$. In the former case,  $e^K|W|^2=0$  and in the latter $e^K|W|^2\not=0$ when
 evaluated at the constant maps $\phi$, respectively.

The moment map condition in (\ref{n3sol}), $\mu^a=0$, together with the remaining constant gauge transformations imply
that the constant maps $\phi$ take values in a K\"ahler quotient of the sigma model target space $S$. It remains to investigate
$D_i W=0$. Suppose that we have chosen some constant maps $\phi=\phi_0$. If $W(\phi_0)=0$, then  $D_i W=0$ implies that
$\partial_i W(\phi_0)=0$. So $W$ and its first derivative vanish at $\phi=\phi_0$.
On the other hand if $W(\phi_0)\not=0$, $D_i W=0$ relates the value of the first derivative of $W$ to that
of the K\"ahler potential at $\phi=\phi_0$.

\newsection{Supersymmetric Quotients}

Supersymmetric solutions of ${\cal N}=1$ four-dimensional supergravity theories can be constructed by
taking quotients of maximally supersymmetric solutions with respect to a discrete subgroup of the isometry group. Here we shall not
investigate all possible cases, instead we shall present an explicit construction
of an  $N=3$ background
from a discrete quotient of $AdS_4$. A similar question has been raised in \cite{FigueroaO'Farrill:2007kb}
in the context of ${\cal{N}}=2$ supergravity theory.
To proceed, consider the gravitino Killing spinor equation equation for an $N=3$ solution
which is locally isometric to $AdS_4$. We take the gauge connection to be trivial and so the scalars to be constant. As $W$ and $K$ are constant, it is convenient to
set
\be
W = -i R e^{i \theta}
\ee
for real $R$, $\theta$, with $R>0$. Furthermore, define $\ell$ by
\be
\ell = {e^{-{K \over 2}} \over R}
\ee
and set
\be
\hat{\epsilon}= e^{-{i\theta \over 2}} \epsilon_L + e^{{i\theta \over 2}} \epsilon_R \ .
\ee
Observe that $\hat{\epsilon}$ is Majorana. Then the Killing spinor equation
implies that
\be
\nabla_\mu {\hat{\epsilon}} +{1 \over 2 \ell} \gamma_\mu {\hat{\epsilon}}=0~.
\ee
The general solution to this equation has been constructed in
\cite{FigueroaO'Farrill:2007kb} using the same notation. In particular, one defines the following
real basis for $AdS_4$:
\bea
{\bf{e}}^0 &=& \ell \cosh \rho  (dt + {1 \over 2} r^2 dx)~,
\cr
{\bf{e}}^1 &=& {\ell \over 2} r^2 \cosh \rho dx~,
\cr
{\bf{e}}^2 &=& \ell d \rho~,
\cr
{\bf{e}}^3 &=& {\ell \over r} \cosh \rho dr \ ,
\eea
for $x, \rho \in \bR$, $t \in [0,2 \pi)$, $r>0$. The smooth quotient
is obtained by making the identification $x \sim x+2k$.
In order to demonstrate how taking this quotient
breaks the supersymmetry from $N=4$ to $N=3$, it
suffices to exhibit four Majorana spinors which are globally
well-defined on $AdS_4$, such that three of these spinors
remain globally well-defined in the quotient geometry, whereas the
fourth fails to be globally well-defined.
These Majorana spinors can be read off directly from equation
(24) of \cite{FigueroaO'Farrill:2007kb}:

\bea
{\hat{\epsilon}}_1 &=& e^{{i \pi \over 4}}
\bigg( 2r (\cosh {\rho \over 2} -i \sinh {\rho \over 2}) (1+e_{12})
+2r (\sinh {\rho \over 2} -i \cosh {\rho \over 2})(e_1-e_2) \bigg)~,
\cr
{\hat{\epsilon}}_2 &=& 2 e^{it} (\cosh {\rho \over 2} + i \sinh {\rho \over 2})1
-2 e^{it} (\sinh {\rho \over 2} +i \cosh {\rho \over 2}) e_2
\cr
&+& 2 e^{-it} (\cosh {\rho \over 2} -i \sinh {\rho \over 2}) e_1
+2 e^{-it} (\sinh {\rho \over 2} -i \cosh {\rho \over 2}) e_{12}~,
\cr
{\hat{\epsilon}}_3 &=& 2i e^{it} (\cosh {\rho \over 2} + i \sinh {\rho \over 2})1
-2i e^{it} (\sinh {\rho \over 2} +i \cosh {\rho \over 2}) e_2
\cr
&-& 2 ie^{-it} (\cosh {\rho \over 2} -i \sinh {\rho \over 2}) e_1
-2i e^{-it} (\sinh {\rho \over 2} -i \cosh {\rho \over 2}) e_{12}~,
\cr
{\hat{\epsilon}}_4 &=& i e^{{i \pi \over 4}}
\bigg( {2 \over r} (1-ir^2 x)(\cosh {\rho \over 2} -i \sinh {\rho \over 2}) 1
-{2 \over r}(1+ir^2 x)(\sinh {\rho \over 2} -i \cosh {\rho \over 2}) e_1
\cr
&-&{2 \over r}(1-ir^2 x)(\sinh {\rho \over 2} -i \cosh {\rho \over 2}) e_2
-{2 \over r}(1+ir^2 x)(\cosh {\rho \over 2}-i \sinh {\rho \over 2}) e_{12} \bigg) \ .
\eea
Clearly, ${\hat{\epsilon}}_1$, ${\hat{\epsilon}}_2$
and ${\hat{\epsilon}}_3$ remain well-defined on making the
identification $x \sim x+2k$. However, as
${\hat{\epsilon}}_4$ contains terms linear in $x$,
${\hat{\epsilon}}_4$ fails to be globally well-defined in this
quotient of $AdS_4$, and hence this solution is an $N=3$ solution. It may worth re-investigating the number of supersymmetries
preserved by this solutions after introducing appropriate flat but no trivial gauge and scalar fluxes.

\newsection{Conclusions}

We have solved the Killing spinor equations of ${\cal N}=1$ supergravity coupled to any number of vector and scalar multiplets.
In particular, we have determined the geometry of spacetime in all cases. We have shown
that there are backgrounds with any number of supersymmetries ranging from $N=1$ to $N=4$.  $N=1$ backgrounds admit
a single null, integrable, Killing vector. $N=2$ backgrounds  admit either a single parallel, null, vector field or three Killing vector
fields. In the former case, the spacetime has an interpretation as a pp-wave.  In the latter, the metric can be written
in special coordinates, and the spacetime is of co-homogeneity one with homogenous section either $\bR^{2,1}$ or $AdS_3$.
Such backgrounds can be thought of as domain walls.
$N=3$ backgrounds are locally maximally supersymmetric. In addition there are backgrounds which admit $N=3$
supersymmetry which can be constructed as discrete identifications of maximally supersymmetric ones. The maximally
supersymmetric backgrounds are locally isometric to either $\bR^{3,1}$ or $AdS_4$.

We have not been able to solve explicitly all the equations. Supersymmetry imposes strong
restrictions in all backgrounds which admit more than one supersymmetry,  $N>1$. Some of the remaining equations are either holomorphic
flow or standard flow type of equations. So many qualitative results can be obtained by investigating the properties of
the vector fields which generate the flow. In particular, the $N=2$ domain wall backgrounds are associated with standard flow equations.
Explicit solutions can be obtained for special models.
Although, we have given an example of an $N=3$ background which can be constructed as discrete identification of
a maximally supersymmetric one based on \cite{FigueroaO'Farrill:2007kb}, we have not investigated all $N=3$ backgrounds that can be obtained in this way.
It may be possible to construct all such backgrounds utilizing the results of \cite{simon}.

\section*{Acknowledgements}
 The work of UG is funded by the Swedish Research Council.

\setcounter{section}{0}

\appendix{Integrability conditions}

There are three integrability conditions that can be derived from the Killing spinor equations in section \ref{killing}. The first is obtained by commuting two gravitino variations,
\bea
\left[ R_{\mu\nu,\rho\sigma}\gamma^{\rho\sigma} + 2(\partial_iK\, {\cal D}_{[\mu}{\cal D}_{\nu]} \phi^i-\partial_{\bar i}K\,{\cal D}_{[\mu}{\cal D}_{\nu]} \phi^{\bar i}) +2 e^{K}W \bar W \gamma_{\mu\nu} \right]\eps_L \notag\\
+ 4i e^{K/2}D_i W {\cal D}_{[\mu} \phi^i \gamma_{\nu]}\eps_R&=&0~,\qquad\qquad
\eea
the second by commuting the gravitino and gaugino variations,
\bea
2\nabla_\mu (F_{\rho\sigma}^a \gamma^{\rho\sigma}-2i\mu^a)\eps_L
- i e^{K/2}W (F_{\rho\sigma}^a \gamma^{\rho\sigma}-2i\mu^a) \gamma_\mu \eps_R = 0~,
\eea
and the third by commuting the gravitino and scalar variations,
\bea
2 ({\cal D}_\mu{\cal D}_\rho \phi^i)\gamma^\rho\eps_R +  e^K G^{i\bar j} (D_{\bar j} \bar W) W \gamma_\mu \eps_R \qquad\qquad \qquad &&\notag\\
\qquad \qquad + {\cal D}_\rho \phi^i \gamma^\rho \left(  (\partial_iK\, {\cal D}_\mu \phi^i-\partial_{\bar i}K\,{\cal D}_\mu \phi^{\bar i})\eps_R + i e^{K/2}\bar W \gamma_\mu \eps_L \right)\notag\\
 + 2i e^{K/2}\left[ \tfrac{1}{2}(\partial_l K\, {\cal D}_\mu \phi^l-\partial_{\bar l}K\,{\cal D}_\mu \phi^{\bar l})G^{i\bar j}D_{\bar j}\bar W \right.\qquad\qquad\notag\\
 \qquad\qquad+  \left. \partial_l G^{i\bar j}{\cal D}_\mu \phi^l D_{\bar j}\bar W  +  \partial_{\bar l} G^{i\bar j}{\cal D}_\mu \phi^{\bar l} D_{\bar j}\bar W +  G^{i\bar j}{\cal D}_\mu D_{\bar j}\bar W \right] \eps_L = 0 ~.
\eea
It is clear from the integrability condition of the gravitino that the holonomy of the supercovariant connection is included in $Pin_c(3,1)$.

\setcounter{section}{0}

\end{document}